\documentclass[a4paper,10pt]{article}

\usepackage{amsmath}
\usepackage{amssymb}
\usepackage{amsthm}
\usepackage{mathtools}
\usepackage{parskip}
\usepackage{graphicx}
\usepackage{psfrag}
\usepackage{subcaption}
\usepackage{url}
\usepackage{hyperref}

\usepackage{xtab,booktabs}

\usepackage{aligned-overset}
\usepackage{authblk}
\usepackage{color}
\usepackage{enumitem}  

\usepackage[hmargin=2.5cm,vmargin=2.5cm]{geometry}

\numberwithin{theo}{section}

\usepackage{xcolor}

\newcommand{\R}{\mathbb R}
\newcommand{\Sp}{\mathbb S}

\newcommand{\Z}{\mathbb Z}

\newcommand{\T}{\mathbb T}

\usepackage{bbm}

\author{%
Thomas Geert de Jong\\
Temple University Japan\\ \texttt{t.g.de.jong.math@gmail.com}
}

\title{Designing learning in high dimensional oscillator networks with low dimensional read-out}

\begin{document}
\maketitle

\begin{quote}
\noindent \textbf{Abstract:} 
In this paper we investigate a oscillator network based reservoir computer with a large number of oscillators and a low dimensional read-out. The read-out is a function on the average phases with respect to each oscillator population. Hence, this read-out provides a robust measurement of the oscillator states. We consider a low number of populations which leads to a low-dimensional read-out.   
Here, the task is time-series prediction. The input time-series is introduced via a forcing term. After a training phase the input is learned. Importantly, the training weights are introduced in the forcing term meaning that the oscillator network is left untouched. Hence, we can apply classical methods for oscillator networks. Here, we consider the continuum limit for Kuramoto oscillators by using the Ott-Antonsen Ansatz.  Consequently, a mean field reservoir computer arises. The success and failure of the reservoir computer is then studied by bifurcations in the coupling and forcing parameter space. We will also show that the average phase read-out can naturally arise when considering the read-out on the phase states. Finally, we give numerical evidence that at least 4 oscillator populations are necessary to learn chaotic target dynamics.

\noindent \textbf{Keywords:} reservoir computer, complex systems, bifurcation theory, dynamical systems
\end{quote}

\section{Introduction}

Due to the widespread adoption of artificial intelligence energy consumption of data centers has rapidly increased~\cite{iea2025,goralski2020artificial}. Hence, efficient computing devices have become a critical research area. 
Designing low-energy computing components that fit into a conventional computing architecture has great merit. 
But it can have its limitations as it forces a process into an architecture~\cite{sebastian2020memory}. Instead we can build a computing architecture around a process to facilitate its deployment.  

A process first approach lies at the core of physical computing~\cite{de2007fundamentals}. In physical computing a physical system is considered with a set of state variables. These state variables evolve following classical or quantum mechanical laws. This is completely different from the instruction-based operations performed on logical bits in conventional computing~\cite{von1993first}. 
The goal in physical computing is then to  design the system in such a way that physical laws perform information-processing by evolving the states. 

The physical computing framework considered in this work is physical reservoir computing \cite{nakajima2020physical}.  From a theoretical perspective reservoir computing can be viewed as a type of recurrent neural network where an input is driven through a massive number of coupled nodes without trainable weights, called the \textit{reservoir}. The reservoir is connected to a single layer of trainable weights which are chosen such that the reservoir's output are mapped into a target. Although originally the reservoir was conceived as a computational operation that is implemented in software it turns out that physical processes, in particular low-energy processes~\cite{vandoorne2014experimental,baltussen2024chemical}, can be used as reservoirs.

In physical reservoir computing measurements are needed to \textit{read-out} the reservoir states as these are fed into the artificial layer with trainable weights. The map that takes the reservoir states into the trainable weight layer is referred to as the read-out. Reading-out physical reservoir states comes with its own set of problems for example measurements can be sensitive to noise or the full state cannot be measured. Hence, we would like to have a robust and practical quantity to read-out from the reservoir. An average reservoir state would be a good candidate. However, if we average over all the states the read-out can be too low-dimensional  to capture nonlinear high-dimensional targets. Multiplexing the states is a method to increase the dimension of the read-out~\cite{timemultiplexing} but this introduces another artificial operation as states from the past must be memorized by the read-out. In this manuscript we simply increase the dimension of the read-out by considering multiple different node populations for which the read-out components are given by the population specific averages.  We note it has been shown that physical node populations given by biological neurons can be used as a physical reservoir computer for classification tasks~
\cite{sumi2023biological}.  

In this work the theoretical description for physical reservoir computing is given by the Omnipresent Computing framework introduced in \cite{dejong2025harnessing}. In omnipresent computing the reservoir is given by a coupled oscillator network subject to input forcing with trainable components only appearing in the forcing term. Hence, it is fundamentally different from Oscillator Neural Networks \cite{todri2024computing,pmlr-v202-keller23a,ricci2021kuranet,miyato2024artificial} where the networks is not forced but trained by considering the pairwise coupling strengths as trainable weights. Hence, Omnipresent Computing is \textit{omnipresent} in the sense that only a general network is needed and not a specific structure that for example has to be obtained through an optimization algorithm.

To simulate a high dimensional oscillator network we consider the continuum equations of a Kuramoto oscillator omnipresent computer in the mean-field limit. More specifically, we consider the so-called Ott-Antonsen Ansatz \cite{ott2008low} to obtain a complex ODE for the average evolution of each oscillator population. It is in this setting that our results are obtained.

Formally, we consider a time-series prediction problem cast within the reservoir framework.  Let
$u: \R \rightarrow \mathcal{M} $ be a smooth function with $\mathcal{M}$ a smooth manifold. Consider the time series $\{ u(0),u( \Delta t),u( 2 \Delta t), \ldots\}$ with $\Delta t>0$. We aim at predicting the next time-step. More specifically, given a finite-length time sequence of $n$-consecutive steps 
we define a reservoir, $R: \mathcal{N} \times \mathcal{M} \rightarrow \mathcal{N}$ and train a $G:   \mathcal{N} \rightarrow \mathcal{M}$ using the available data such that the composition $\hat{G}: r \mapsto G(\hat{R}(r))$ with  $\hat{R}: r \mapsto R(r, G(r))$ approximates $u$. More explicitly, denote by $\hat{G}^{(k)}= \underbrace{\hat{G} \circ \ldots \circ \hat{G}}_{k{\rm - times}}$ then we aim to find a $G$ such that
\begin{gather}
\begin{aligned}
\hat{G}^{(k)}(u( n \Delta t )) \approx u( (k+n) \Delta t). \label{eq:mainproblem}
\end{aligned}
\end{gather}
Hence, the LHS of \eqref{eq:mainproblem} will be referred to as the prediction of $u$. The operator $G$ will be the composition of a linear trainable weight matrix with a fixed read-out function. In our setting $R$ will be induced by the flow of the oscillator network. Our study will then consider the dynamics of $R,\hat{R}, G, \hat{G}$ for specific $u$, as we vary the parameters of the oscillator network and choose appropriate read-outs. In particular, number of oscillator populations is a parameter. These populations are engineered by selecting a distribution for the natural frequency as well as an input component which drives the population.

In this work we consider a wide range of topics for high dimensional oscillator networks with low dimensional read-out. The topics are themed around the dynamics induced by changing the oscillator networks' parameters. The results will be presented by using toy models. Let us give an overview:

\begin{itemize}
\item[-] Rigorous analysis of Continuum Limit (CL) oscillator networks: We can perform a full parameter analysis for 1-dimensional linear input which tells us a priori when the omnipresent computer is successful. 
\item[-] Usage of symmetries to reduce parameter study of CL oscillator networks: For input dimension 2 it becomes challenging to perform analysis. However, we can impose symmetries on the input and oscillator states that allow us to reduce the state space. 
\item[-] Naturally arising average phase read-out for Finite Dimensional (FD) oscillator networks: In the CL settings we use the average phase in the read-out. We show that for an FD oscillator network with non-linear 2-dimensional target input the prediction with a component-wise read-out on the phases, in other words a high dimensional read-out, approximates the population specific averaged phase, a low dimensional read-out.
\item[-] Achieving successful predictions of chaotic time series by CL networks through increasing oscillator populations: In the previous settings we consider the oscillator population equal to the dimension of the target input. However, for 3-dimensional chaotic systems it turns out that this is insufficient. By engineering more than 3 populations the CL oscillator network can sustain chaotic dynamics.  
\end{itemize}
This paper is organized as follows: In Section \ref{sec:over} the Omnipresent Computing framework for Kuramoto oscillators is revised which will be referred to as the FD network, and then the CL network is derived, in Section \ref{sec:results} the main results are presented and in Section \ref{sec:conc} we discuss how the results fit in with respect to the broader field and propose future research directions.

\section{Omnipresent Computing: Reservoir computing by oscillator networks \label{sec:over}}

We briefly revise the general set-up for reservoir computing \cite{jaeger2001echo,jaeger2004harnessing}. We define an input $u: \R \rightarrow \mathcal{M}$ where $\mathcal{M}$ is an $M$-dimensional manifold.  We consider the dynamical system obtained by the flow of the ordinary differential equation

\begin{align}
 \frac{dr}{dt} = R(r,u) ,  \label{eq:res_main}
\end{align}
where $r(t) \in \mathcal{N}$ is an $N$-dimensional manifold and $R: \mathcal{N} \times \mathcal{M} \rightarrow  \mathcal{N}$. 

\begin{figure}
\centering
\includegraphics[width=13cm]{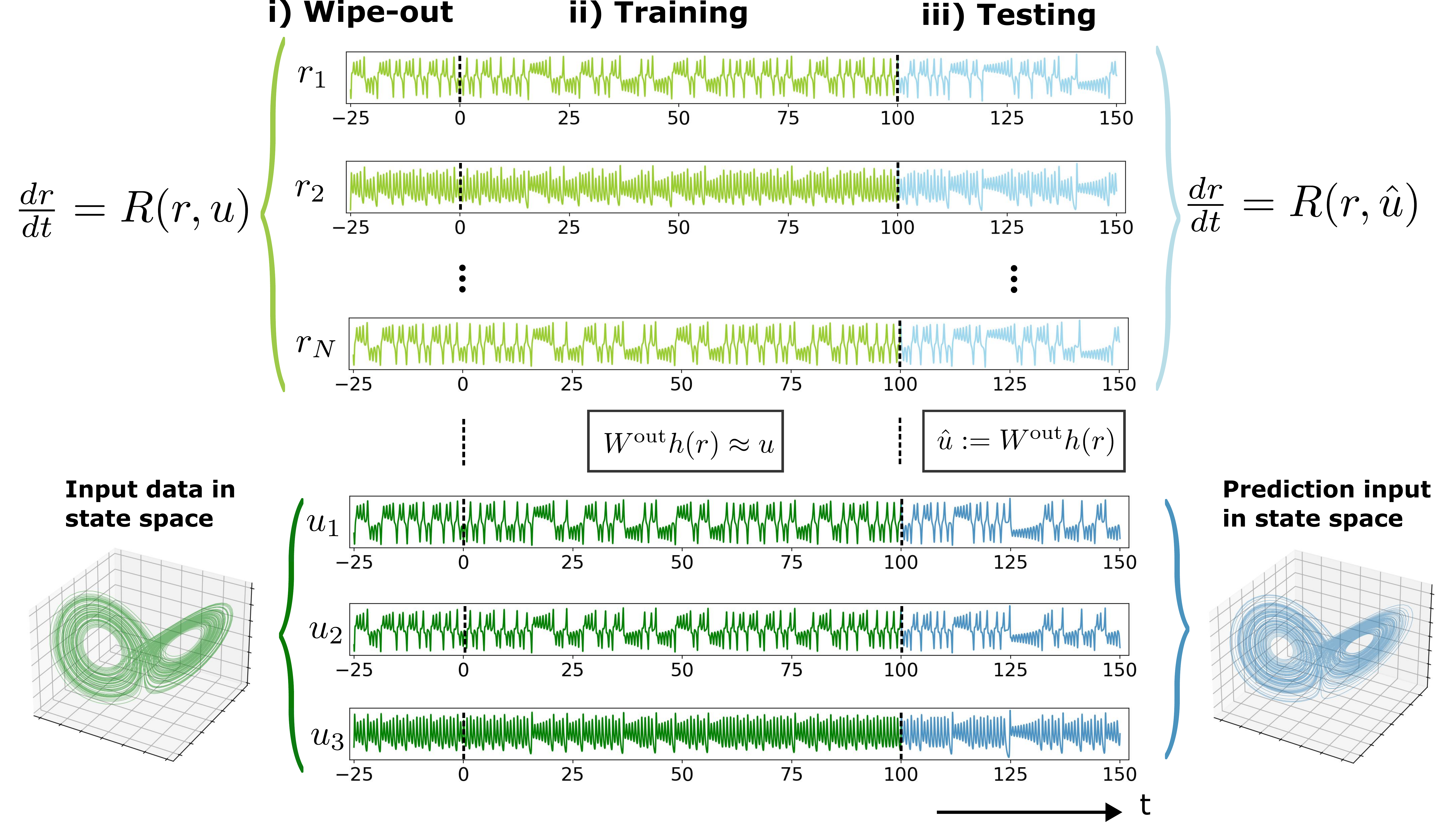}
\caption{\textbf{Overview deployment autonomous reservoir computer:} The reservoir evolution is described by a $u$-driven ODE. Here the input is given by a solution of Lorenz system.  i) During wipe-out transients are removed by evolving the $u$-driven reservoir. ii) During training the $u$-driven reservoir is also evolved but a linear weight matrix, $W^{\rm out}$ is trained so that the read out of the states, $h(r)$, is fitted to $u$.  iii) During testing the $u$ is substituted by $W^{\rm out}h(r)$ so that the resulting system can predict $u$ autonomously. \label{fig:rc_overview}}
\end{figure}

The scheme to constructing and evaluating a reservoir consist of three steps: 
\begin{itemize}
\item[1.] \textbf{Wipe-out:} We consider the initial value problem corresponding to \eqref{eq:res_main} by setting $r(-T_{\rm wipe}) =r_0$ and compute the solution $r$ for all $t \in [-T_{\rm wipe},0)$ as to remove transients (Fig. \ref{fig:rc_overview}i).
\item[2.] \textbf{Training:}   We consider a yet to be defined $h: \mathcal{N} \rightarrow \mathcal{N}_{\rm ro}$, referred to as read-out function and compute a linear operator  $W^{\rm out}:\mathcal{N}_{\rm ro} \rightarrow \mathcal{M}$ such that $W^{\rm out} h  \approx u$ on $t \in [0, T_{\rm train})$ for an appropriate error measure  (Fig. \ref{fig:rc_overview}ii). The linear operator $W^{\rm out}$ is referred to as the weight matrix.
\item[3.] \textbf{Testing:} In \eqref{eq:gov} we substitute $u$ by $W^{\rm out} h(r)$. Let us denote the new dependent variable by $\hat{r}$. Then, the solution with initial value ${r}(T_{\rm train}) =\hat{r}(T_{\rm train})$ will be used to determine performance of the reservoir for $t \in [T_{\rm train}, T_{\rm train}+T_{\rm test})$  (Fig. \ref{fig:rc_overview}iii). The performance will be evaluated by using error measures and by comparing Lyapunov exponents of $u$ to $r$. 
\end{itemize}
Observe that we have cast the time series prediction problem \eqref{eq:mainproblem} in a continuous setting.

\subsection{Finite Dimensional (FD) oscillator network}

In \cite{dejong2025harnessing} a framework for harnessing omnipresent oscillator networks as computational resource is presented. We consider the setting in the context of a Kuramoto oscillator network. We consider $N$ oscillators represented by $\theta \in \T^N:=\underbrace{\Sp \times \Sp \times \ldots \times \Sp}_{N{\rm -times}}  $ where  $\Sp := \R / 2 \pi \Z$. Now \eqref{eq:res_main} takes the form:
\begin{align}
\frac{d \theta_{k}}{dt} =  \omega_{k,j} +  \underbrace{ {\frac{K}{N} \sum_{i=1}^N [\sin ( \theta_i - \theta_{k} ) ]}}_{\rm coupling} + \underbrace{ F \sin (  u_{\ell_j} - \theta_{k})}_{\rm forcing} \qquad \forall k \in \mathcal{I}_{j}, \; \; j=1,2, \ldots P\label{eq:gov}
\end{align}
where $K \geq 0$ is referred to as the coupling constant, $F \geq 0$ is referred to as the forcing constant, $\ell_{j} \in \{1,2, \ldots, M\}$,  $\mathcal{I}_{j}$ is referred to as the $j$th population set which is an index set that contains the indexes $k$ of $\theta$ which are forced by $u_{\ell_j}$ and have natural frequencies  $\omega_{k,j}$ sampled from the $j$th Cauchy distribution $g_{j}(\omega)$ given by
\[
g_{j}(\omega) = \frac{\Delta_j}{\pi ( (\omega - \omega_{0j})^2 + \Delta_j^2) },
\]
$P$ is the number of population sets. We suppose that $\mathcal{I}_{j}$ all have the same size and define $N_{\rm in }:= |\mathcal{I}_{\ell}|$ (Fig.~\ref{fig:osc_overview}i). We note that the coupling in \eqref{eq:gov} is all-to-all. We refer to the reservoir corresponding to \eqref{eq:gov} as Finite Dimensional (FD) Kuramoto reservoir. We note that for $K=0$ and $P=1$ Equation \eqref{eq:gov} becomes the classical Kuramoto model~
\cite{kuramoto1975self}.\\

To analyze the dynamics associated to the oscillators in $\mathcal{I}_{\ell}$ we can introduce the $j$th population complex order parameter: 
\begin{align*}
z_j  = \frac{1}{N_{\rm in}} \sum_{k \in \mathcal{I}_j} e^{i \theta_k}. 
\end{align*}
The complex order parameter $z_j$ corresponds to the centroid of the phases and describes the collective rhythm produced by the oscillators \cite{kuramoto1975self,strogatz2000kuramoto}. Let $\rho_j, \psi_j$ be given by $z_j = \rho_j e^{i \psi_j }$. Then, $\rho_j$ determines the phase coherence of the oscillators. For $\rho_j$ close to 0 the oscillators spread out over the circle and for $\rho_j$ close to 1 the oscillators are clumped together. Finally, $\psi_j$ determines the average phase.

\subsection{Continuum Limit (CL) oscillator network}

We now consider a continuum limit of \eqref{eq:gov} such that we can reduce the resulting governing equations to an ODE with dependent variable $z_j$.  This continuum limit relies on the so-called Ott-Antonsen Ansatz \cite{ott2008low}.

\begin{figure}[ht]
\centering
\includegraphics[width=15cm]{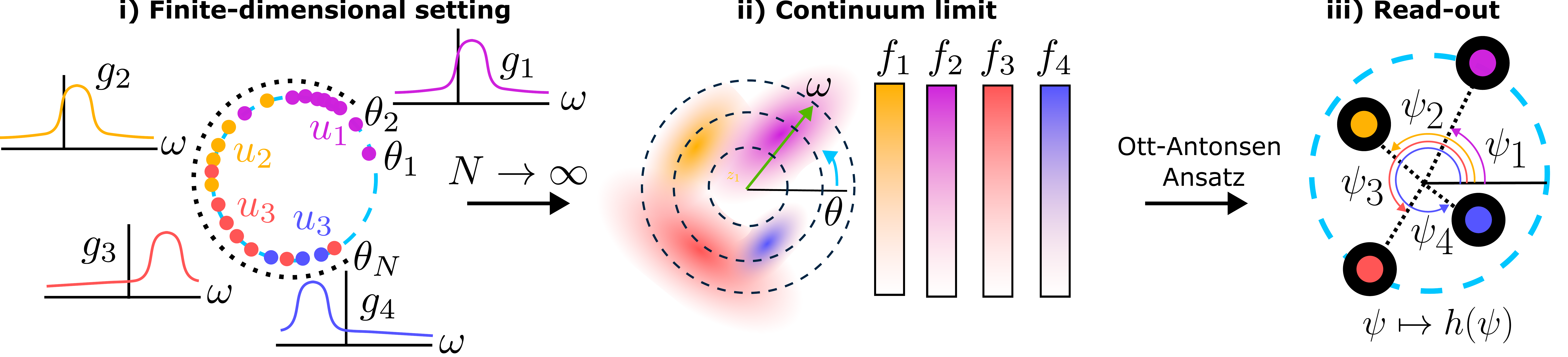}
\caption{\textbf{Overview example for relation between FD oscillator network and CL oscillator network:} i) For the FD oscillator network we consider $M=3$ and $P=4$ with $\ell = (1,2,3,3)$. Note that the  populations driven by $u_3$ are not identical since $g_3,g_4$ are different. ii)  In an infinite dimensional setting we can describe the oscillator populations by density functions over the phases and natural frequencies. iii) Using the Ott-Antonsen Ansatz we reduce our study to the average population dynamics. Specifically, for CL oscillator networks we consider the read-out as a function over the average phases, $\psi$.}
\label{fig:osc_overview}
\end{figure}

The density function $f_j(\theta, \omega, t)$ is defined such that at a time $t$, the fraction of oscillators with natural frequencies in $g_j(\omega)$ and coupled to $u_{\ell_j}$ has phases between $\theta$ and $\theta + d\theta$ and natural frequencies between $\omega$ and $\omega+ d\omega$ is given by $f_j(\theta, \omega, t) d \theta d\omega$ (Fig.~\ref{fig:osc_overview}ii). This yields that 
\[
\int_{-\infty}^{\infty} \int_{0}^{2\pi} f_j(\theta, \omega ,t) d\theta  d \omega = 1, \qquad  \int_{0}^{2\pi}f_j(\theta, \omega ,t) d\theta  = g_j(\omega).
\]

We then consider the continuity equations
\begin{align}
\frac{d f_j}{dt} + \frac{\partial v_j f_j}{\partial \theta} =0, \label{eq:ct2}
\end{align}
where
\[
v_j = \omega + K  \int_{-\infty}^{\infty} \int_{0}^{2\pi} \sin(\theta' - \theta) \left( \sum_{k=1}^M f_k(\theta', \omega ,t) \right) d \theta' d \omega + F \sin(u_{\ell_j}(t) -  \theta).
\]
The complex order parameter corresponding to the $j$th density function is given by
\[
z_j(t) = \int_{-\infty}^{\infty}\int_{0}^{2 \pi} e^{i \theta} f_j(\theta, \omega,t) d\theta d\omega.
\]
We write $f_j$ as a Fourier series:
\[
f_j(\theta, \omega ,t) = \frac{g_j(\omega)}{2 \pi} \left(  1 + \sum_{n=1}^{\infty} \left[ f_j^{(n)}(\omega,t) {e}^{i n \theta } + f^{(n)}_j(\omega,t)^* {e}^{-i n \theta }  \right] \right).
\]
We consider the Ott-Antonsen Ansatz: $f_j^{(n)}(\omega,t) = \alpha_j(\omega,t)^n$ with  $\alpha_j(\omega,t)$ that can be analytically continued from the real $\omega$-axis into the lower half of the complex $\omega$-plane for all $t \geq 0$, $|\alpha_j(\omega,t)| \rightarrow 0$ as ${\rm Im}(\omega) \rightarrow -\infty$,  $|\alpha_j(\omega,0)| \leq 1$ for real $\omega$. The governing equations for the reservoir \eqref{eq:res_main} in the dependent variables $\rho_j, \psi_j$ given by $z_j = \rho_j e^{i \psi_j}$ take the form:
\begin{gather}   
\begin{aligned}
    \dot{\rho}_j &= - \Delta_j \rho_j + \frac{K}{2M}   (1 - \rho_j^2) \sum_{k=1}^M \left[ \rho_k \cos(\psi_j - \psi_k) \right] + \frac{F}{2} (1- \rho_j^2) \cos(\psi_j - u_{\ell_j}), \\
    \dot{\psi}_j &= \omega_{0j} - \frac{K}{2M} \frac{1 + \rho_j^2}{\rho_j} \sum_{k=1}^M \left[ \rho_k \sin(\psi_j-\psi_k) \right] - \frac{F}{2} \frac{1+\rho_j^2}{\rho_j} \sin(\psi_j - u_{\ell_j}).
\end{aligned} \label{eq:rho_psi}
\end{gather}

Details of derivations can be found in Appendix \ref{app:deriv_ct}. We refer to the corresponding reservoir as Continuum Limit (CL) Kuramoto reservoir.

We will fix $\Delta_j = \Delta_1$. Now define $(\hat{t},\hat{F},\hat{K},\hat{\omega}_{0j}) = ({t},{F},{K},{\omega_{0j}})/ \Delta_1$. Then, in the hat-parameters and hat-variable $\Delta_1$ cancels out. Therefore, we will drop the hat and consider $\Delta_1=1$. 

For the tasks in the next section it appears that $\rho$ is not necessary for the readout, $h$. Hence, we will consider $h (\psi)$ (Fig.~\ref{fig:osc_overview}iii). We will see that for analysis purposes Equation \eqref{eq:rho_psi} is convenient to work with. We note that standard numerical solvers struggle with Equation \eqref{eq:rho_psi}. It is more convenient to apply the solver to the complex ODE corresponding to $z_j$ for which the expression can be found in Appendix \ref{app:deriv_ct}.

\begingroup
 \begin{table}[ht]
\centering
\caption{Summary of parameters}
\label{tab:over}

 \begin{tabular}{ll}
\toprule
Symbol & Description  \\
\midrule
$u$ & time-series function for prediction task \\
$\hat{u}$ & $W^{\rm out} h(\psi)$ for CL network, $W^{\rm out} h(\theta)$ for FD network \\ 
$N$ & number of oscillators \\
$M$ & dimension of time-series $u$ \\
$\mathcal{I}_{j}$ & $j$th population index set \\
$P$ & number of oscillator populations \\
 $\ell_j$ & index for the $u$-component that forces oscillators in $\mathcal{I}_{j}$ \\
$\omega_{0j}$ & location parameter Cauchy distribution corresponding to oscillators in $\mathcal{I}_{j}$  \\
$h$ & read-out function from the states \\
$K$ & coupling constant \\
$F$ & forcing constant \\
$T_{\rm wipe}$ & wipe-out duration\\
$T_{\rm train}$ & training duration\\
$T_{\rm test}$ & testing duration \\
 \bottomrule
\end{tabular}
\end{table}
\endgroup

\section{Results \label{sec:results}}

We present results for prediction tasks of an $M$-dimensional time-series $u$ for $P$-populations of oscillators. Although the main goal is the study of the CL-reservoir we also provide results in connection to the FD-reservoir. 

\subsection{CL-reservoir for $P=M=1$: asymptotic stability of predictions \label{sec:res1d}}

We consider \eqref{eq:rho_psi} with a linear input $u(t)=ct$. 

\subsubsection{Reservoir configuration \label{sec:1dconfig}}

We will not consider a wipe-out phase. We take $h(\psi) =[1, \psi]$. If during training there exist an open set of initial conditions for which the solution $(\rho,\psi)$ satisfies $(\rho,\psi-u)(t) \rightarrow (\rho_0,\psi_0)$ for $t \rightarrow \infty$ then $W^{\rm out} = [-\psi_0, 1]$ or more explicitly, $\hat{u}=\psi - \psi_0$.

\subsubsection{Bifurcation analysis \label{sec:1dbifu}}

We will perform a full bifurcation analysis during both the training and testing phase and obtain when the reservoir converges asymptotically to a successful prediction of $u$.

Let $\phi: = \psi - ct$. Define $\Omega := c - \omega_0 $.  Then, the governing equations during training are given by
\begin{gather}
\begin{aligned}
\dot \rho &=  -  \rho + \frac{K}{2}\rho (1-\rho^2) + \frac{F}{2}(1-\rho^2) \cos (\phi), \\
\dot \phi &=  -\Omega - \frac{F}{2} \left( \rho + \frac{1}{\rho} \right) \sin (\phi).
\end{aligned} \label{eq:wipeout1d}
\end{gather}
We observe that we are interested in the attracting fixed points of \eqref{eq:wipeout1d} by the conditions imposed in Section \ref{sec:1dconfig}. We note that Equation \eqref{eq:wipeout1d} is equivalent to the setting considered in \cite{childs2008stability, antonsen2008external}. However, the authors in \cite{childs2008stability,antonsen2008external} fix $K$  and vary $\Omega$. But the overall analysis carries over to our setting. 

\begin{figure}[ht]
\centering
\includegraphics[width=16.5cm]{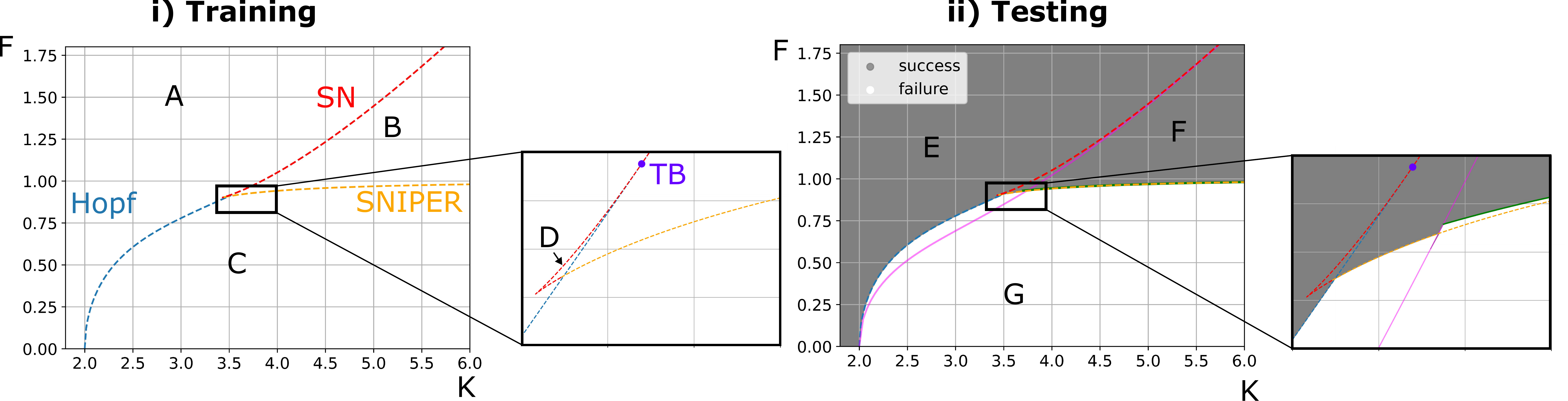}
\caption{
\textbf{Bifurcation diagrams for $\Omega=1$:} i) In the  training phase the Hopf, Saddle-Node, SNIPER bifurcation divide the parameter domain into the subdomains A,B,C,D. In A,B,D training is successful. ii) In the testing phase the green and the magenta curve subdivide the parameter domain into the subdomains E,F,G. In E,F testing is successful if training is successful. Hence, the prediction is successful for $(A \cup B \cup D) \cap (E \cup F)$.}  
\label{fig:bifu_1d}
\end{figure}

We will fix $\Omega=1$ and perform a bifurcation analysis for the dynamics during training (Fig.~\ref{fig:bifu_1d}i). The domain is subdivided into the subdomains A,B,C,D by a Hopf, Saddle-Node and SNIPER bifurcation. We note that since there is a Takens-Bogadanov point there is also a homoclinic bifurcation but since it does not influence the stability of the fixed points it will not be relevant to our study and therefore has been omitted. In domain A,B there is 1 attracting fixed point and in D there are 2 attracting fixed points. In domain C there is only a repelling fixed point enclosed by an attracting limit cycle. Hence, only in A,B,D training can be successful.

Let's continue with the testing phase. The governing equations during testing are
\begin{gather}
\begin{aligned}
\dot \rho &=  -  \rho + \frac{K}{2}\rho (1-\rho^2) + \frac{F}{2}(1-\rho^2) \cos (\psi_0), \\
\dot \phi &= - \Omega - \frac{F}{2} \left( \rho + \frac{1}{\rho} \right) \sin (\psi_0).
\end{aligned} \label{eq:test1d}
\end{gather}
Observe that the vector field is independent of $\phi$. Hence, we are only interested in the stability of the $\rho$-equation. Denote the $\rho$-derivative of the RHS of the $\rho$-equation by $J_\rho$.  Observe that we can assume that $\psi_0 \in \T$. Take $\psi_0 \neq \pi/2,3\pi/2$. Then, using that $\dot \rho =0, \dot \phi =0, J_\rho =0$ and $\sin(\phi)^2 + \cos(\phi)^2 =1$ we obtain
\begingroup\makeatletter\def\f@size{5}\check@mathfonts
\def\maketag@@@#1{\hbox{\m@th\large\normalfont#1}}%
\begin{align}
\tiny
F= \frac{\sqrt{K} \sqrt{2 K^2 \left(\Omega ^2+47\right)+2 K^4+43 K^3+K \left(7 \Omega ^2+52\right)+2 \left(\Omega ^2+4\right) +  \sqrt{1+4K}\left(-7 K^3-38 K^2-3 K \left(\Omega ^2+12\right)-2 \left(\Omega ^2+4\right)\right)}}{\sqrt{ 1 + 6 K + 9 K^2 + 2 K^3 - \sqrt{1+4K}\left(3 K^2+4 K+1 \right)}} \label{eq:magenta_curve}
\end{align}\endgroup
Observe that the numerator is zero for $K=2$ and that the numerator is increasing in $\Omega$ if $K >2$. 
Additionally, the denominator is greater than zero if $K \geq 2$. The curve has been colored in magenta (Fig.~\ref{fig:bifu_1d}ii).  

Setting $\psi_0 = 3 \pi/2$ we arrive at
\begin{align}
F= \frac{\sqrt{K-2} \sqrt{K}}{K-1} \Omega \label{eq:green_curve}
\end{align}
The curves given by \eqref{eq:magenta_curve} and \eqref{eq:green_curve} have a unique intersection for $K\geq 2$ given by
\begin{align}
\Omega = \frac{\sqrt{(K-2) (K-1)^2 \left(2 K^3+\left(17-5 \sqrt{4 K+1}\right) K^2-4 \left(\sqrt{4 K+1}+1\right) K+4 \left(\sqrt{4 K+1}-1\right)\right)}}{\sqrt{2} \sqrt{K^3-6 K^2+10 K-4}}. \label{eq:KOmega}
\end{align}
Observe that \eqref{eq:KOmega} is increasing in $K$. Suppose $\Omega$ is fixed. Then denote by $K_{\Omega}$ the $K$ satisfying \eqref{eq:KOmega}. Then, for $K< K_{\Omega}$ we have that $\dot{\rho}=0, \dot{\phi}=0, J_{\rho}<0$. Hence, only $K> K_{\Omega}$ is of interest, which is the green curve (Fig.~\ref{fig:bifu_1d}).

The magenta and green curve enclose 3 domains: E,F,G (Fig.~\ref{fig:bifu_1d}). Given a fixed point of \eqref{eq:wipeout1d} with non-zero eigenvalues  we can determine how the stability changes when considering \eqref{eq:test1d}. In E all fixed points of \eqref{eq:wipeout1d} become attracting for \eqref{eq:test1d}. In F saddle and attracting fixed points of \eqref{eq:wipeout1d} become attracting for  \eqref{eq:test1d}, while repelling fixed points of \eqref{eq:wipeout1d} remain repelling for \eqref{eq:test1d}. In G all fixed points of \eqref{eq:wipeout1d} become repelling for \eqref{eq:test1d}. 

Finally, putting the training and testing results together we find that in the domain $(A \cup B \cup D) \cap (E \cup F)$ the prediction is successful (Fig.~\ref{fig:bifu_2d}). Notably in the region $B \cap G$ training is successful but testing fails which is a phenomenon familiar to the full field of neural networks. Another region of note is $D \cap E$ in which a provable untrained attractor exists \cite{o2025confabulation}.\\

Although we have performed a full analysis, we thought it still insightful to visualize the parameter dependent solutions numerically, \textbf{Movie I, II}. We note that the biases are computed by 
$\tilde{\psi}_0 = \tilde{\psi}(T_{\rm train})-u(T_{\rm train})$ where the tildes indicate the numerically obtained values. Additionally, instead of visualizing the states in $(\rho,\phi)$ we visualized the states in $\hat{z}:= \rho e^{i \phi}$.

\subsection{CL-reservoir for $P=M=2$: stability of prediction up to small oscillations \label{sec:2dcl}}

Let $u= (u_1,u_2)$, $\ell_1=1,\ell_2=2$,  We reduce the dimension of \eqref{eq:rho_psi} by introducing symmetries: $u_1=-u_2$, $\omega_{01} = -\omega_{02}$, $\psi_1 = -\psi_2,\rho_1 = \rho_2$. 

Now in terms of dependent variables we only have to consider $\rho_1, \psi_1$. Consequently, let us drop the subscripts.  From \eqref{eq:rho_psi} we now obtain:
\begin{gather}
\begin{aligned}
\dot{\rho} &=  - \rho  + \frac{K}{4}(1 - \rho^2) \left( \rho + \rho \cos(2 \psi) \right)  +     \frac{F}{2} (1 - \rho^2) \cos(\psi - u) , \\
\dot{\psi} &=  \omega_{0} - \frac{K(1+\rho^2)}{4} \sin(2 \psi ) - \frac{F}{2} \frac{(1+ \rho^2)}{\rho} \sin (\psi - u).
\end{aligned} \label{eq:gov_m2}
\end{gather}
Finally, we let $u(t)= ct$.  

Observe that the governing equations are non-autonomous. Unfortunately, we cannot transform the system to a 2-dimensional autonomous ODE. Therefore, our analysis will be lead by the numerics. We will fix the parameters: $\omega_0 =1, c = 2$.

\subsubsection{Reservoir configuration}

We initialized the reservoir with a wipe-out. For training we define the read-out function as $h(\psi) = [1, \psi]$. Observe that this is the same form as the read-out considered in Section \ref{sec:res1d}. The trainable weight matrix $W^{\rm out} \in \R^2$ is then obtained using ridge regression \cite{hoerl1970ridge}. Since $\psi \in \Sp$ we need to consider its lift, see Appendix \ref{app:ridge}. The exact parameter configuration for the CL-reservoir can be found in Appendix \ref{app:param}

\subsubsection{Bifurcation diagrams}

\begin{figure}[ht]
\centering
\includegraphics[width=16cm]{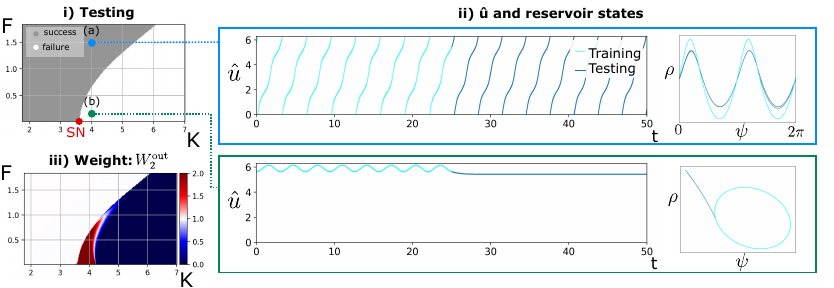}
\caption{\textbf{CL-reservoir for symmetric linear $u$:} i) The intersection of the boundary between success and failure with the line $F=0$ coincides with a saddle node bifurcation.  ii) For successful parameter choices the prediction exhibits small oscillations around the target.  The limiting dynamics in the state space for successful prediction is characterized by  a stable periodic orbit while for failed prediction the solution is attracted to a fixed point.  iii) For successful parameter choices we observe that $W_2^{\rm out} \approx 1$.}
\label{fig:bifu_2d}
\end{figure}

We will consider the reservoir in the $(F,K)$-plane. During training and testing we measure the short-time performance with the Normalized Mean Square Error (NMSE), Appendix \ref{app:nmse}.  The NMSE is either small or large, which corresponds to success or failure of the reservoir (Fig.~\ref{fig:bifu_2d}i). Even when the NMSE is small there are small oscillations present in the prediction (Fig.~\ref{fig:bifu_2d}ii). 
Hence, the dynamics is qualitatively different from the setting with 1-dimensional input. We note that these oscillations are also present in $\rho, \psi$ during the training and testing phase.\\

For $F=0$ \eqref{eq:gov_m2} becomes autonomous. We can then show analytically that at the boundary of the two domains a saddle-node bifurcation occurs (Fig.~\ref{fig:bifu_2d}i). More specifically, for $K< K_{SN}$ the system has no fixed points and for $K> K_{SN}$ there is an attracting fixed point  with a saddle (with the saddle only existing up until some $K$ value), see Appendix \ref{app:Fis0}. Recall that in the 1-dimensional input case the fixed point analysis was performed on moving co-ordinates.  In this setting if the solution of \eqref{eq:gov_m2} converges to a fixed point this implies failure of prediction. Now suppose $F>0$ and we are in the testing phase with $W^{\rm out}$ given, then a straightforward analysis can be applied as the equations reduce to an autonomous 2-dimensional ODE. 

For successful prediction we find that during training $W^{\rm out}_2 \approx 1$ (Fig.~\ref{fig:bifu_2d}iii). This means that the prediction is given by $\hat{u} \approx W^{\rm out}_1 + \psi$. Hence, upto a constant $\psi$ captures the dynamics of $u$. It will then come as no surprise that success of prediction corresponds to the solution converging to a stable periodic orbit enclosing the origin (Fig.~\ref{fig:bifu_2d}ii). Failure of prediction corresponds to the solution  converging to a stable fixed point (Fig.~\ref{fig:bifu_2d}ii).

\subsection{FD-reservoir for $P=M=2$: natural occurrence of CL-reservoir predictions \label{sec:2dfd}}

In the continuum limit of \eqref{eq:gov} the order parameters naturally arise as a means to study the dynamics. In the previous sections we observed that the angular order parameter, $\psi$, can be used in the read-out function to obtain successful predictions for suitable parameters. That brings us to the question whether the angular order parameter naturally arises in the finite dimensional setting if we would consider a read-out function over the states of the individual oscillators $\theta_i$. 

We now consider 2-dimensional periodic input: $u_1(t) = \sin(t)/10, u_2(t) = \cos(t)/10$. Additionally, we assume $\omega_{01}=\omega_{02}$. Hence, the oscillator populations are identified by the components of $u$.

\subsubsection{Reservoir configuration}

At $t=0$ the oscillators will be placed equidistantly over the circle. The oscillators with odd and even index will be forced by $u_1$ and $u_2$, respectively. The order parameter is only expected to give us insight into the dynamics if the number of oscillators is sufficiently large. Hence, we will take $N=2000$. 

For training we use the read-out function $h(\theta) = [1,\sin(\theta)]$. We note that if we would consider quadratic terms in $\sin(\theta)$ the majority of weights corresponding to $\sin^2(\theta)$ are close to zero. The linear weight matrix $W^{\rm out} \in \R^{2 \times (N+1) }$ obtained during training is computed using ridge regression.

It will be important to choose a small $\Delta_1$ since values in the tail of the Cauchy distribution will turn out to negatively effect our result. Using a distribution with less flat tails, such as a normal distribution, would also resolve this problem.

For the exact parameter configuration we refer to Appendix \ref{app:param}.

\subsubsection{Connection between reservoir states, weights and $\psi$}

\begin{figure}[ht]
\centering
\includegraphics[width=16cm]{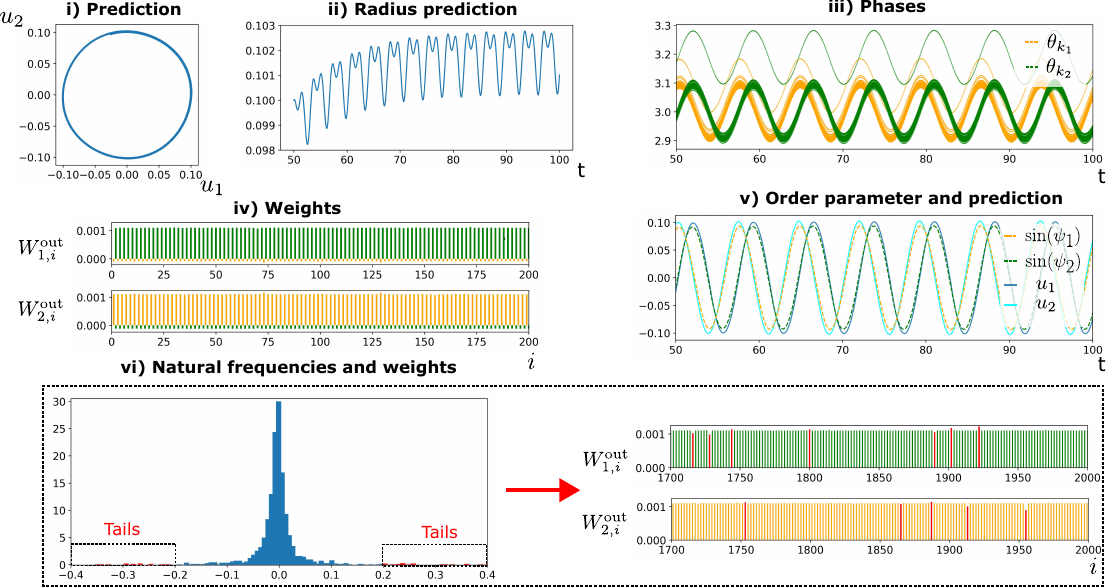}
\caption{\textbf{FD-RC for periodic motion on a circle in the plane:} i,ii) The prediction is successful but exhibits small oscillations which appear to be bounded over time.  iii) Oscillator populations have the same frequency but are translated in time differently in correspondence with their population, i.e., $\theta_{k_i}$ with $k_i \in \mathcal{I}_i$ for $i=1,2$.  iv) Weight matrix excluding the bias term exhibits regular periodic pattern with respect to indexes. v) The function  $\sin(\psi_j)$ approximate $u_i$. vi) Natural frequencies in the tails give rise to weights that deviate significantly from the mean. } 
\label{fig:finite}
\end{figure}

We can successfully configure the oscillator network for the prediction task (Fig.~\ref{fig:finite}i)
. During testing we find that the radius of circle parametrized by the input, $\sqrt{\hat{u}_1^2+\hat{u}_2^2}$, exhibits small oscillations which indicates that the network cannot perfectly capture the input but only approximates it (Fig.~\ref{fig:finite}ii). Furthermore, the average dynamics of the radius is an increasing and converging function meaning that a bounded error gets introduced as the system switches from training to testing.

Exploring the oscillator states we observe that  the oscillators all have approximately the same frequency. However, the oscillator states differ by how they are translated in the spatial and time direction. We will ignore the spatial direction and only consider the translations in the time direction. We observe that  oscillators in the same population, $\theta_{k_i}$ with $k_i \in \mathcal{I}_i$ for $i=1,2$, are translated the same (Fig.~\ref{fig:finite}iii). This grouping also appears in the weight matrix,  $W^{\rm out}$.   Observe that  $W^{\rm out}_{1,i}$ are used in the prediction of $\hat{u}_1$ and $W^{\rm out}_{2,i}$ are used in $\hat{u}_2$. We omit the bias terms from $W^{\rm out}_{1,i}$, $W^{\rm out}_{2,i}$. Then, we observe that $W^{\rm out}_{1,i}$ and $W^{\rm out}_{1,i}$ are approximately constant for $i \in \mathcal{I}_1$ and $i \in \mathcal{I}_2$ (Fig.~\ref{fig:finite}iv). But this implies that
\begin{align}
\hat{u}_1 \approx \frac{1}{N_{\rm in}} \sum_{k  \in \mathcal{I}_2}\sin(\theta_{k}) =: \sin(\psi_2), \qquad \hat{u}_2 \approx \frac{1}{N_{\rm in}}\sum_{k \in \mathcal{I}_1}\sin(\theta_{k}) =: \sin(\psi_1). \label{eq:up_con}
\end{align}
In other words, the average phase naturally arises in this setting (Fig.~\ref{fig:finite}v). We do not have an explanation why the reservoir uses $\psi_2$ to predict $u_1$ and $\psi_1$ to predict $u_2$. For different choices of parameters this phenomenon persists. 

The $\omega_i$ are sampled from a Cauchy-distribution. Cauchy-distributions have long tails. We observe that the weights corresponding to the components in the tails have weights with much higher variance than the weights not in the tails (Fig.~\ref{fig:finite}vi). More specifically, we consider $W^{\rm out}_{1,i}$ with $i \in \mathcal{I}_2$ ,  $W^{\rm out}_{2,i}$ with $i \in \mathcal{I}_1$, which are the weights furthest away from zero. Hence, these tails have a negative effect on maintaining the equality in \eqref{eq:up_con}. 

We note that the regular ``$W^{\rm out}$-pattern'' (Fig.~\ref{fig:finite}iv), is not necessary for successful prediction of the input. Hence, the prediction that can be expressed in $\psi$ describes a special class of solutions.

\subsection{CL-reservoir for $M=3$: learning chaotic time-series \label{sec:3d}}
We will take $u=(x,y,z)$ to be given by a solution on the chaotic attractor of the Lorenz system:
\begin{gather}
\begin{aligned}
\frac{dx}{dt} &= \sigma (y - x), \\
\frac{dy}{dt} &= x(\rho - z) - y,\\
\frac{dz}{dt} &= x y - \beta z, 
 \label{eq:lorenz}
\end{aligned}
\end{gather}
with the parameters $\sigma=10, \beta=8/3,\rho=28$~\cite{lorenz1963deterministic}.

\subsubsection{Reservoir configuration}

We consider the read-out function $h(\rho, \psi) = [1, \sin(\psi), \sin(\psi)^2] \in \R^{1+2P}$. It provides a straightforward way to map from the oscillator topology, $\Sp$, into the components of the target time-series, $\R$. 

Full details on the reservoir configurations can be found in Appendix \ref{app:param}

\subsubsection{$P\geq 4$ can sustain chaos}

\begin{figure}[ht]
\centering
\includegraphics[width=13cm]{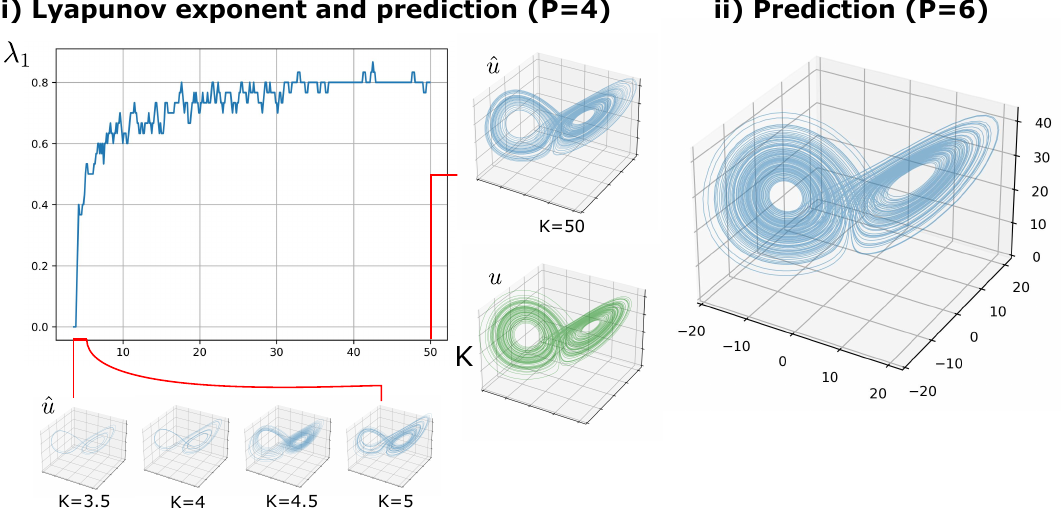}
\caption{\textbf{CL-RC for chaotic input:} i) We consider 4 oscillator populations and graph the evolution of the oscillator's leading Lyapunov exponent versus $K$ with fixed $F$. We note that we only consider the first two decimals of the Lyapunov exponent. Between $K=3.5$ and $K=5$ a period doubling bifurcation occurs which leads to the chaotic attractor. It appears that the leading Lyapunov exponent is bounded well below the leading Lyapunov exponent of the Lorenz attractor. ii) When we consider 6 populations we can find parameters for which the prediction resembles the Lorenz attractor and has leading Lyapunov exponent close to the leading Lyapunov exponent of the Lorenz attractor.}
\label{fig:M4M6}
\end{figure}

For $P=1$ the CL-RC cannot exhibit chaotic dynamics and therefore, cannot be successful at the learning task. Numerically, we couldn't succeed for $P=2,3$. However, for $P=4$ we found that the system was able to sustain chaotic dynamics during testing. We considered $(\ell_1,\ell_2,\ell_3,\ell_4)=(1,2,3,3)$, meaning that one population is forced by the $x,y$-component of the Lorenz system and that two populations are forced by the $z$-component of the Lorenz system. As we increase $K$ there is a critical parameter where the system undergoes a period doubling bifurcation and the leading Lyapunov exponent becomes greater than zero resulting in a Lorenz-like attractor (Fig.~\ref{fig:M4M6}i). Varying the parameters we were unable to find a chaotic attractor that has leading Lyapunov exponent close to the leading Lyapunov exponent of the Lorenz attractor. Additionally, upon visual inspection we observe that the wings of the predicted attractor are deformed in comparison to the Lorenz attractor. 

Now let's take $P=6$ with $(\ell_1,\ell_2,\ell_3,\ell_4,\ell_5,\ell_6)=(1,1,2,2,3,3)$. Hence, 2 oscillator populations are forced by each component of the Lorenz system.  We obtain attractors which are more similar to the Lorenz attractor (Fig.~\ref{fig:M4M6}ii). Indeed, visually it is hard to tell the predicted attractor apart from the Lorenz attractor. Additionally, the leading Lyapunov exponent is now given by 0.90 which is much closer to the leading Lyapunov exponent of the Lorenz attractor which is approximately 0.906. 

\section{Concluding remarks and future work \label{sec:conc}}

In this work we investigated forced high dimensional oscillator networks with low dimensional read-out using the Ott-Antonsen Ansatz. Via a dynamics study we investigated how parameters of the system determine its performance for time-series prediction tasks. Specifically, we saw that by varying $K,F,P$ we can achieve the target dynamics. Additionally, we observed that the low dimensional read-out on $\psi$ can naturally emerge when considering a read-out on the oscillator phases.

Due to the Ott-Antonsen Ansatz we could consider relatively low-dimensional differential equations. As a result we could perform detailed investigations of the dynamics. However, by using Ott-Antonsen Ansatz we are also removing many solutions from our analysis which might improve the performance of the reservoir. A numerical analysis on the relation of reservoir performance to Ott-Antonsen Ansatz would be meaningful. Additionally, there is strong numerical evidence that the network structure influences the performance of the reservoir which cannot be approached with the analysis of this work since all-to-all connectivity is required.

We note that there is a rich literature on Kuramoto based computing. A large portion of these computing frameworks are Kuramoto inspired in the sense that the differential equation structure of oscillator networks is not present and hence the analysis techniques from the Kuramoto literature do not carry over. In Oscillator Neural Networks (ONN)~\cite{todri2024computing} a oscillator network structure is present. However, in ONN the input is typically encoded  by the natural  frequencies or in-phase between oscillators and the trainable weights are the pairwise oscillator coupling strength. For FD oscillator networks  the oscillator network is left untouched during training and testing which allows us to perform the analysis in this work. Additionally, it appears that ONN are typically not used for memorization of dynamics. Finally, from a more general perspective this specific work differs from other oscillator based computing frameworks by approximating the target input using only a low dimensional variable.  Hence, the  theoretical approximation approaches as in~\cite{HART_genct} were not pursued. 

In this work we showed that for the Omnipresent Computing framework the reservoir computing component is sufficiently vanilla that oscillator network theory can be directly applied. This in turn gave us information about the performance of the reservoir computer. Hence, making an interdisciplinary bridge. Due to this  vanilla reservoir computing component it is expected that a wide variety of results can be found at this new intersection.

\textbf{Acknowledgments:} This work was supported by JST CREST grant JPMJCR2014. During this research Thomas de Jong was also affiliated to Kanazawa University and Duke-Kunshan University. We would like to thank Eddie Nijholt (University of Sao Paolo) and Edmilson Roque dos Santos (Max Planck Institute PKS) for their helpful comments during the revision stages.

\appendix

\section*{Appendix}

\section{Derivation governing equations CL oscillator network \label{app:deriv_ct}}

We continue from \eqref{eq:ct2}. The velocity can be written as
\begin{align}
v_j = \omega + \frac{K }{2Mi} \left(  \sum_{k=1}^M z_k e^{-i \theta }- \sum_{k=1}^M \left[z_k \right]^* e^{i \theta } \right) + \frac{F}{2i} \left( e^{i( u_{\ell_j} -  \theta )} - e^{-i( u_{\ell_j} -  \theta) } \right) \label{eq:v2}
\end{align}
We write $f_j$ as a Fourier series:
\[
f_j(\theta, \omega ,t) = \frac{g_j(\omega)}{2 \pi} \left(  1 + \sum_{n=1}^{\infty} \left[ f_j^{(n)}(\omega,t) {e}^{i n \theta } + f^{(n)}_j(\omega,t)^* {e}^{-i n \theta }  \right] \right)
\]
Substituting \eqref{eq:v2} into \eqref{eq:ct2} and making the ansatz $f_j^{(n)}(\omega,t) = \alpha_j(\omega,t)^n$ we obtain the following ODE:
\[
\frac{d \alpha_j}{dt} = \frac{F}{2}\left( e^{-i u_{\ell_j} } - e^{i u_{\ell_j} } \alpha_j^2   \right) - i \omega \alpha_j - \frac{K}{2M}\left( \alpha^2_j \sum_{k=1}^M z_k - \sum_{k=1}^M \left[z_k\right]^* \right).
\]
Observe that 
\[
z^*_j = \int^{\infty}_{-\infty} \alpha_j(\omega,t) g_j(\omega) d\omega.
\]

Then integrating with respect to $\omega$ and taking the complex conjugate we have an expression in terms of the complex order parameter:
\[
\frac{dz_j}{dt} = \frac{F}{2} \left(  e^{i u_{\ell_j} } -  e^{-i u_{\ell_j}  } z_j^2 \right) + (i \omega_{0j} - \Delta_j) z_j - \frac{K}{2M}\left(z^2_j \sum_{k=1}^M \left[z_k\right]^* - \sum_{k=1}^M z_k\right)
\]
Note that we have assumed that $\alpha_j(\omega,t)$ can be analytically continued from the real $\omega$-axis into the lower half of the complex $\omega$-plane for all $t \geq 0$ and that $|\alpha_j(\omega,t)| \rightarrow 0$ as ${\rm Im}(\omega) \rightarrow -\infty$ and that $|\alpha_j(\omega,0)| \leq 1$ for real $\omega$.

%
%
%


\section{Technical details of reservoirs}

Code is available on my \href{https://github.com/mathowl/mean_field_reservoir_computing}{github repo}.

In Table \ref{tab:over} we give an overview of symbols specific to this section.

\begingroup
 \begin{table}[ht]
 \caption{Summary of symbols}
\label{tab:symbols}
\centering
 \begin{tabular}{ll}
\toprule
Symbol & Description \\
\midrule
$\Delta t$ & time-step numerical solver  \\
$\beta$ & regularization parameter \\
$ n_{\rm wipe}  $ & number of wipe-out steps  \\
$ n_{\rm train}  $ & number of training steps  \\
$ n_{\rm test}  $ & number of testing steps  \\

 \bottomrule
\end{tabular}
\end{table}
\endgroup

\subsection{Parameter configurations of reservoirs \label{app:param}}
The parameters of the reservoirs are presented in Table \ref{tab:param_A} and Table \ref{tab:param_B}.

\begingroup
 \begin{table}[!ht]
\centering
\caption{Summary of parameters for reservoirs from  Section \ref{sec:2dcl}, Section \ref{sec:2dfd}  \label{tab:param_A}}
 \begin{tabular}{l|llll}
\toprule
& \multicolumn{2}{c}{Reservoir} \\
\hline
Symbol &  CL$(P,M)=(2,2)$(Section \ref{sec:2dcl}) & FD$(P,M)=(2,2)$ (Section \ref{sec:2dfd})  \\
\hline
$T_{\rm wipe}$ &  25 &  25 \\
 $T_{\rm train}$ & 25 & 25 \\
 $T_{\rm test}$ &  25 & 50 \\
 $\beta$  &  $10^{-5}$  & $10^{-4}$ \\
 $\Delta t$  & $10^{-3}$ & $ 5 \cdot 10^{-3}$\\
 $\Delta$ &   1  & 0.01\\
 $ N $ &  NA & 2000 \\ 
 $\omega_{0i}$ & 1 & 0  \\
 \bottomrule
\end{tabular}
\end{table}
\endgroup

\begingroup
 \begin{table}[!ht]
\centering
\caption{Summary of parameters for reservoirs in Section \ref{sec:3d}}
\label{tab:param_B}
 \begin{tabular}{l|llll}
\toprule
& \multicolumn{2}{c}{Reservoir} \\
\hline
Symbol &  CL$(P,M)=(4,3)$ (Section \ref{sec:3d}) & CL$(P,M)=(6,3)$(Section \ref{sec:3d})   \\
\hline
$T_{\rm wipe}$ &  50 &  50 \\
 $T_{\rm train}$ & 200 & 200 \\
 $T_{\rm test}$ &  100 & 100 \\
 $\beta$  &  $10^{-11}$  & $10^{-11}$ \\
 $\Delta t$  & $10^{-2}$ & $ 10^{-2}$\\
 $\Delta$ &   1  & 1\\
 $ F $ &  55 & 55 \\
  $ K $ &  NA & 80 \\
  $c $ &  0.01  & 0.01 \\
 $\omega_{0i}$ & (0.43, -0.07, -0.09,  0.03)  & (-0.14, -0.11,  0.4 , -0.34, -0.18, -0.03)  \\
 \bottomrule
\end{tabular}
\end{table}
\endgroup

    



\subsection{Numerics of training: Ridge regression on $\Sp$ \label{app:ridge}}

We proceed by example and present the training scheme for Section \ref{sec:2dcl}.  
 
 Let $n_{\rm train}=  \left\lfloor  T_{\rm train}/\Delta t \right\rfloor $. We represent the time-series vector obtained during training  for $\psi$ and $u$ by the $n_{\rm train}$-dimensional vectors $\Psi \in \R^{n_{\rm train}}$ and $U \in \R^{n_{\rm train}}$,  respectively.   
\begin{align*}
W^{\rm out} = \mathop{\arg \min}\limits_{W} \|  W G(\hat{h}(\Psi))  - G(U) \|_2^2  +  \beta \| W \|_2^2,
\end{align*}
with $\hat{h}: \T^{n_{\rm train}} \rightarrow \T^{2 \times n_{\rm train} }$  given by applying $h$ on the spatial vectors at each time-step, where $\beta > 0$ is the regularization parameter and $G: \T^{2 \times n_{\rm train} } \rightarrow \R^{2 \times n_{\rm train}}$ the natural lift. The $W^{\rm out}$ can be then be computed using ridge regression:
\[
W^{\rm out} =   (G(\hat{h}(\Psi)) G(\hat{h}(\Psi))^T  + \beta I)^{-1} G(\hat{h}(\Psi)) U.
\]

\subsection{Error measure: Normalized Mean Square Error (NMSE) \label{app:nmse}}

We consider an error on the time-discretization of the prediction. Let $n_{\rm test} = \left\lfloor T_{\rm test}/\Delta t \right\rfloor $. The Normalized Mean Square Error (NMSE) is defined by 
\begin{align*}
{\rm NMSE} :=  \frac{1}{M} \sum_{j=1}^M \frac{\sum_{i=1}^{n_{\rm test}} | u_j(t_i)-\hat{u}_j(t_i) |^2}{ \sum_{i=1}^{n_{\rm test}} | {u}_j(t_i) |^2 } , \qquad t_i := T_{\rm train} + T_{\rm test}+ (i-1) \Delta t.
\end{align*}
This defined the testing error. For the training error we use an analogous definition.

\subsection{Computation acceleration}
All the computations were performed on a CPU. When appropriate the computations were accelerated following the directives of \cite{de2023mlncp}.

\section{Supporting bifurcation analysis for Section \ref{sec:2dcl}}

\subsection{Saddle-Node bifurcation $F=0$ \label{app:Fis0}}

We will consider \eqref{eq:gov_m2} for general $\omega_0$. Although $\rho=0$ can now be studied we restrict to $\rho \in (0,1]$ in accordance with $F>0$ setting.

To find the saddle-node bifurcation we set $\dot \rho =0, \dot \psi =0$ and ${\rm det}(J)=0$, where $J$ denotes the Jacobian. Using $\dot \rho =0, \dot \psi =0$, $\sin^2(2\psi)+\cos^2(2\psi)=1$ and solving for $K$ we obtain  
\begin{align}
K = -\frac{2 \left(\rho ^4 \omega^2_0-2 \rho ^2 \omega^2_0+\rho ^4+2 \rho ^2+\omega_0 ^2+1\right)}{\left(1-\rho ^2\right) \left(\rho ^2+1\right)^2} \label{eq:Keqd2}
\end{align}
under the assumptions that
\begin{align}
\left| \frac{-4 + K - K \rho^2}{K (-1 + \rho^2)}  \right| \leq 1, \qquad \frac{4 \omega_0}{K (1 + \rho^2)} \leq 1. \label{eq:sntechr}  
\end{align}
Substituting $K$ in ${\rm det}(J)$ by \eqref{eq:Keqd2} we obtain that the following equality is satisfied:

\begin{align*}
K = \frac{8}{\rho ^4-4 \rho ^2+3}, \qquad
\omega_0 =  \sqrt{\frac{\left(\rho ^2+1\right)^3}{\left(3-\rho ^2\right) \left(\rho ^2-1\right)^2}}
\end{align*}

Let $K>\frac{8}{3}$. Substituting the above equalities in \eqref{eq:sntechr} we find that they are satisfied. Then, we parametrize $\omega_0$ in terms of $K$:
\begin{align*}
\omega_0= \sqrt{\frac{\left(\sqrt{K (K+8)}-3 K \right)^3}{8 K \left(\sqrt{K (K+8)}-K\right)}}
\end{align*}

Specifically, for $\omega_0 =1$  we find that $K_{\rm SN} \approx 3.61$.

\subsection{Testing bifurcations \label{app:supp_bifu}}

\begin{figure}[h]
    \centering
    \begin{subfigure}[b]{0.45\textwidth}
        \centering
        \includegraphics[width=\textwidth]{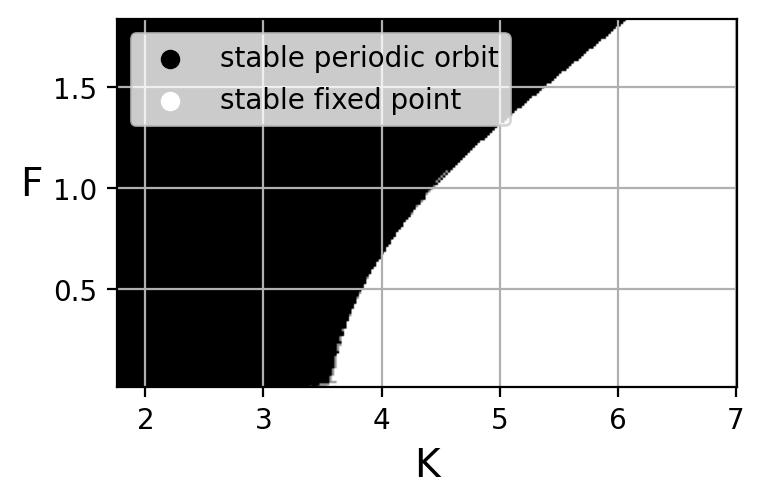}
        \caption{Limiting dynamics in state space during testing}
        \label{fig:first}
    \end{subfigure}
    \hfill
    \begin{subfigure}[b]{0.45\textwidth}
        \centering
        \includegraphics[width=\textwidth]{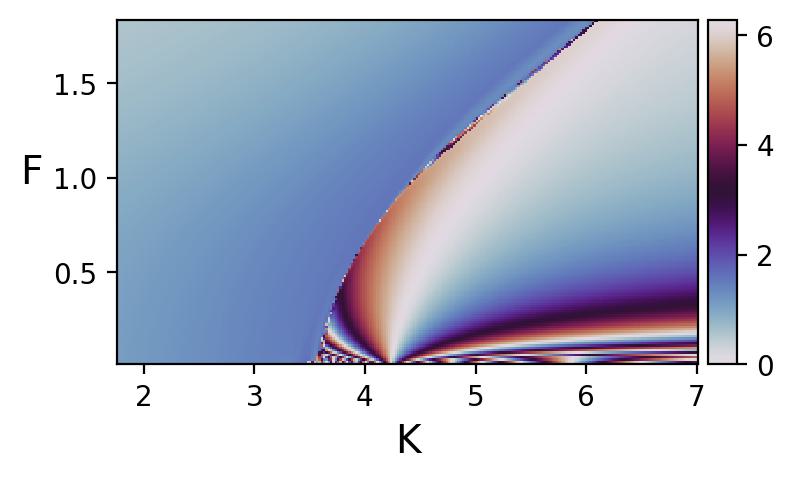}
        \caption{Weights $W_1^{\rm out}$}
        \label{fig:second}
    \end{subfigure}

    \caption{Supporting bifurcation diagrams}
    \label{fig:both}
\end{figure}

We provide supporting results on the limiting dynamics of the solutions during testing (Fig.~\ref{fig:first}). In the domain where testing is successful the solution is attracted to a stable periodic orbit enclosing the origin and in the domain where testing fails the solution is attracted to a stable fixed point. The sudden change in NMSE resulting in the bifurcation diagram for testing, (Fig.~\ref{fig:bifu_2d}), also occurs as a sudden change in $W^{\rm out}_1$ (Fig.~\ref{fig:second}).

\bibliographystyle{alpha}
\bibliography{new_bib.bib}{}

\end{document}